\documentclass[twocolumn,superscriptaddress,showpacs,aps,amsmath,amssymb]{revtex4}
\usepackage{epsfig}
\usepackage{amsbsy}
\usepackage{amsmath}
\usepackage{amsfonts}
\usepackage{graphicx}
\usepackage{latexsym}

\begin{document}

\title{Non-Markovian dynamics of a microcavity coupled to a waveguide in photonic crystals}
\author{Meng-Hsiu Wu}
\affiliation{Department of Physics and Center for Quantum
information Science, National Cheng Kung University, Tainan 70101,
Taiwan}
\author{Chan U Lei}
\affiliation{Department of Physics and Center for Quantum
information Science, National Cheng Kung University, Tainan 70101,
Taiwan}
\author{Wei-Min Zhang}
\email{wzhang@mail.ncku.edu.tw} \affiliation{Department of Physics
and Center for Quantum information Science, National Cheng Kung
University, Tainan 70101, Taiwan}
\author{Heng-Na Xiong}
\affiliation{Zhejiang Institute of Modern Physics and Department of
Physics, Zhejiang University, Hangzhou, People's Republic of China}

\date{June 1, 2010}

\begin{abstract}
In this paper, the non-Markovian dynamics of a microcavity coupled
to a  waveguide in photonic crystals is studied based on Fano-type
tight binding model. Using the exact master equation, we solve
analytically and numerically the temporal evolution of the cavity
coherent state and the associated physical observables.  A critical
transition is revealed when the coupling increase between the cavity
and the waveguide. In particular, the cavity field becomes
dissipationless when the coupling strength goes beyond a critical
value, as a manifestation of strong non-Markovian memory effect. The
result also indicates that the cavity can maintain in a coherent
state with arbitrary small number of photons when it strongly
couples to the waveguide at very low temperature. These properties
can be measured experimentally through the photon current flowing
over the waveguide in photonic crystals.
\end{abstract}

\keywords{Quantum optics, Microcavity, Non-Markovian Dynamics,
Decoherence}

\pacs{42.50.-p, 03.65.Yz, 42.79.Gn}

\maketitle

\section{INTRODUCTION}

Optical microcavities confine light in the micro- and submicro-scale
volumes by resonant recirculation with very high quality factors
\cite{highQPCcavity,highQSOI}. Micro- and submicro-scale volume
ensures that resonant frequencies are more sparsely distributed in
the size-dependent resonant frequency spectrum. Prototypical
micrcavities include the Fabry-Perot microcavities, the silica-based
microdisk, microsphere, and microtoroid whispering gallery cavities,
and the photonic crystal cavities. Devices based on these
microcavities are already received tremendous attentions for a wide
range of applications, including strong coupling cavity QED
\cite{SEcontrol,laserOsiQD,controlQED}, low threshold lasers
\cite{lowthresholdlaser}, biochemical detectors \cite{biosensor,
chemicaldet}, as well as optical traps \cite{PCtrap}. In this paper,
we shall study the non-Markovian dynamics of a microcavity coupled
to a  waveguide in photonic crystals.

A microcavity in photonic crystals is a point defect created in
photonic crystals as a resonator.  Its frequency can easily be tuned
to any value within the band gap by changing the size or the shape
of the defect and therefore can be used to enhance the efficiency of
lasers. While, a waveguide in photonic crystals consists of a linear
defects in which light propagates due to the coupling of the
adjacent defects. By changing the modes of the resonators and the
coupling configuration, the transmission properties of the waveguide
can be manipulated. The most promising application of waveguide is
to control the group velocity, thus potential to application in
storing and buffering light by coupling to a microcavity
\cite{slowlight, largegv, PCCROWfilter}. While, the coupling between
the microcavity and the waveguide is controllable \cite{cavity with
CROW-1}, which can induce non-Markovian dissipation and decoherence
phenomena \cite{cavity with CROW-2}. The non-Markovian dynamics is
an important factor in the practical applications of quantum
information and quantum computation in terms of photons. Therefore,
we shall use the exact master equation we developed recently to
investigate the non-Markovian dynamics of the microcavity field
coupled to a  waveguide in different coupling regime, to explore
possible new applications of microcavities in quantum optics.

The paper is organized as follow. In Sec.~II, we introduce the exact
master equation we developed recently for the reduced density
operator of a cavity coupled to the waveguide as a reservoir. In
this section, the reduced density operator as well as the temporal
evolution of the cavity mode amplitude and the photon number are
derived analytically. The photon current in waveguide is also
calculated directly from the time-dependence of the photon number in
the cavity. In Sec.~III, exact non-Markovian dynamics of the cavity
field is demonstrated numerically through the temporal evolution of
the cavity mode amplitude as well as the photon number inside
cavity. By varying the coupling between the cavity and the waveguide
in different coupled configuration, the significant non-Markovian
memory effect is revealed. Finally, summary and discussion are given
in Sec.~IV.

\section{The microcavity dynamics coupled to a waveguide}

\subsection{Fano-type tight-binding model for a microcavity coupled to a waveguide}

\begin{figure}
\centering
\includegraphics[scale=0.3]{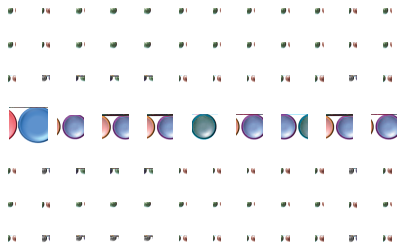}
\caption{(Color online) A schematic plot of a microcavity coupled to
a waveguide in photonic crystals.} \label{mw}
\end{figure}

We consider a microcavity with a single mode coupled to a
 waveguide in photonic crystals, see a schematic plot
in Fig.~\ref{mw}. The microcavity is a point defect created in
photonical crystals as a resonator. While the waveguide consists of
a linear defects in which light propagates due to the coupling of
the adjacent defects. Therefore, the Hamiltonian of the system can
be expressed as a tight binding model:
\begin{align}
H = & \omega_{c} a^{\dag}a +
    \sum_n \omega_{0} a_{n}^{\dag} a_{n} \notag
    \\ & -  \sum_n \xi_{0} (a_{n}^{\dag} a_{n+1} +
{\rm H.c.}) + \xi(a^{\dag }a_{1} + {\rm H.c.}) \ . \label{hh}
\end{align}
Here we have set $\hbar=1$. The first term in Eq.~(\ref{hh}) is the
Hamiltonian of the microcavity in which $a^{\dag},a$ are the
creation and annihilation operators of the single mode cavity field,
with frequency $\omega_{c}$ which can easily be tuned to any value
within the band gap by changing the size or the shape of the defect.
The second and third terms are the Hamiltonian of the waveguide
where $a_{n}^{\dag},a_{n}$ are the photonic creation and
annihilation operators of the resonator at site $n$ of the waveguide
with an identical frequency $\omega_{0}$, and $\xi_{0}$ is the
hopping rate between adjacent resonator modes. Both $\omega_0$ and
$\xi_0$ are experimentally turnable. The last term is the coupling
between the microcavity and the waveguide with the coupling constant
$\xi$. While, the coupling between the cavity and the waveguide is
also controllable by changing the geometrical parameters of the
defect cavity and the distance between the cavity and the waveguide
\cite{cavity with CROW-1}.

The about system of a miscrocavity coupled to a waveguide in
photonic crystals can also be implemented with different type of
micro-resonators, such as Fabry-Perot microcavitties and micro-ring
resonators,  with different coupling and confinement mechanism
\cite{CROWproposal}. However, the dispersion relation of different
kinds of microcavities and micro-waveguides are very similar, it is
only characterized by the free spectral range, the quality factor of
the resonators and the coupling between the resonators
\cite{CROWdelayline}. Therefore Eq.~(\ref{hh}) describes indeed a
large class of a microcavity coupled to a micro-waveguide.
Furthermore, the Hamiltonian (\ref{hh}) can be re-expressed as a
Fano-type model of a localized resonance interacting with continuums
\cite{Fano611866, FanoAnderson-2}:
\begin{align}
H = \omega_{c} a^{\dag} a + \sum_k  \omega_k a_{k}^{\dag} a_{k} +
\sum_k \big[V_k a_{c}^{\dag} a_{k} + {\rm H.c.}\big] \ ,
\end{align}
where $0 \leq k \leq \pi$, $\omega_k$ and $V_k$ are given by:
\begin{align}
\omega_k =  \omega_{0}-2\xi_{0}\cos(k), \ V_k=\sqrt{2/\pi}~\xi
\sin(k), \label{vk}
\end{align}
and $a_{k}^{\dag},a_{k}$ are the creation and annihilation operators
of the corresponding Bloch modes of the waveguide, which is defined
as follow:
\begin{equation}
a_{k}=\sqrt{2 \over \pi} \sum_{n=1}^{\infty} \sin(nk)a_{n} \ .
\end{equation}
After transform the Hamiltonian into this form, we can use the exact
master equation we developed recently to examine the non-Markovian
dynamics of the microcavity coupled to the waveguide in photonic
crystal quantum optics.

\subsection{Exact master equation}

The master equation for the cavity field is given in terms of the
reduced density operator which is defined from the density operator
of the total system by tracing over entirely the environmental
degrees of freedom: $\rho(t) \equiv {\rm tr}_{\rm R} \rho_{\rm
tot}(t)$, where the total density operator is governed by the
quantum Liouville equation $\rho_{\rm tot}(t)= e^{-i H(t-t_0)}
\rho_{\rm tot}(t_0) e^{i H(t-t_0)}$. By integrating over all the
environmental degrees of freedom, based on the Feynman-Vernon
influence functional approach \cite{Fey63118} in the framework of
coherent state path-integral representation \cite{Zhang9062867}, we
obtain the exact master equation for the reduced density operator
\cite{Tu08235311, An07042127, Xiong1005}:
\begin{align}
\dot{\rho}& \left( t\right) =-i\omega'_c(t) \left[ a^{\dag}a,\rho(t)
\right] \notag \\ & +\kappa \left( t\right) \big\{
2a\rho(t) a^{\dag}-a^{\dag}a\rho(t) -\rho(t) a^{\dag}a \big\}  \notag \\
&+\widetilde{\kappa}(t) \big\{ a^{\dag}\rho(t) a+a\rho(t)
a^{\dag}-a^{\dag}a\rho(t) -\rho(t) aa^{\dag} \big\} ,  \label{eme}
\end{align}%
where the time-dependent coefficient $\omega'_c (t)$ is the
renormalized frequency of the cavity, while $\kappa(t)$ and
$\widetilde{\kappa}(t)$ describe the dissipation and noise to the
cavity field due to the coupling with the reservoir. These
coefficients are non-perturbatively determined by the following
relations:
\begin{subequations}
\label{coefs}
\begin{align}
& \omega'_c (t) = -{\rm Im}[\dot{u}(t)u^{-1}(t)] , \label{coefs-a}\\
& \kappa \left( t\right) = -{\rm Re}[\dot{u}(t)u^{-1}(t)],  \label{coefs-b} \\
& \widetilde{\kappa}(t) = \dot{v}(t)-2 v(t){\rm Re}[\dot{u}(t)u^{-1}(t)],  \label{coefs-c}
\end{align}
\end{subequations}
and $u\left( t\right) $ and $v(t)$ satisfy the integrodifferential
equations of motion:
\begin{subequations}
\label{uv-eq}
\begin{align}
& \dot{u}(\tau) +i \omega_c u (\tau)+ \int_{t_0}^{\tau }  d\tau' g
(\tau -\tau') u (\tau') =0,  \label{u-e} \\
&v\left( t\right) =\int_{t_{0}}^{t}d\tau _{1}\int_{t_{0}}^{t}d\tau
_{2}\text{ }\overline{u}\left( \tau _{1}\right)\widetilde{g}\left(
\tau _{1}-\tau _{2}\right) \overline{u}^{\ast }\left( \tau
_{2}\right) ,  \label{v-e}
\end{align}
\end{subequations}
subjected to the initial condition $u (t_0) = 1$ while
$\overline{u}(\tau)\equiv u(t+t_0-\tau)$.

Note that the integral kernels in the above equations are the time
correlation functions of the waveguide: $g(\tau-\tau')$ and $
\widetilde{g}(\tau-\tau')$. These two time-correlation functions
characterize all the non-Markovian memory structures between the
cavity and the waveguide. By defining the spectral density of the
waveguide: $J(\omega) =2\pi \sum_k |V_k|^{2} \delta (\omega
-\omega_k)$, the time-correlation functions are explicitly given by
\begin{subequations}
\label{correla}
\begin{align}
g( \tau -\tau') &=\int_{0}^{\infty }\frac{d\omega}{2\pi} J\left(
\omega \right) e^{-i\omega ( \tau -\tau')} ,   \\
\widetilde{g}(\tau -\tau') &=\int_{0}^{\infty }\frac{d\omega}{2\pi} J(
\omega) \overline{n}(\omega ,T)e^{-i\omega ( \tau -\tau')} ,
\end{align}
\end{subequations}
where $\bar{n}(\omega ,T) = \frac{1}{e^{\hbar \omega /k_{B}T} -1}$
is the average number distribution of the waveguide thermal
excitation at the initial time $t_0$. With the spectrum of the
photonic crystal (\ref{vk}), the spectral density becomes
$J(\omega)=2\pi g(\omega)|V(\omega)|^2$, and $g(\omega)$ is the
density of state:
\begin{align}
& g(\omega)= \frac{dk}{d\omega}=\frac{1}{
\sqrt{4\xi_0^2-(\omega-\omega_0)^2}}, \notag \\
& V(\omega)=\frac{1}{\sqrt{2\pi}}\Big(
\frac{\xi}{\xi_0}\Big)\sqrt{4\xi_0^2-(\omega-\omega_0)^2}.
\label{vks}
\end{align}
Then the spectral density can be explicitly written as
\begin{equation}
\label{eq:J} J(\omega)= \Big(\frac{\xi}{\xi_{0}}\Big)^{2} \sqrt{4
\xi_{0}^{2} -(\omega - \omega_{0})^{2}} \ ,
\end{equation}
with $\omega_{0} - 2 \xi_{0} < \omega < \omega_{0} + 2 \xi_{0}$. In
practical, $\xi_0 \ll \omega_0$, namely the waveguide has a very
narrow band.

The master equation (\ref{eme}) is exact, far beyond the BM
approximation widely used for conventional optical cavities. The
back-reaction effect between the system and environment is fully
taken into account by the time-dependent coefficients, $\omega'_0
(t) $, $\kappa (t) $ and $\widetilde{\kappa}(t) $, in the master
equation (\ref{eme}) through the integrodifferential equations
(\ref{uv-eq}). Thus, the non-Markovian memory structure is
non-perturbatively built into the integral kernels in these
equations.  The expression of the integrodifferential equation
(\ref{uv-eq}) shows that $u(t)$ is just the propagating function of
the cavity field (the retarded Green function in nonequilibrium
Green function theory \cite{Kad62}), and $v(t)$ is the corresponding
correlation (Green) function, as we will see next. Therefore, the
exact master equation (\ref{eme}) depicts the full nonequilibrium
dynamics of the cavity system as well as the waveguide.

\subsection{Exact solutions of the microcavity dynamics}

The main physical observables for the microcavity are the temporal
evolution of the cavity mode amplitude and the photon number inside
the cavity. The cavity mode amplitude is defined by $\langle a(t)
\rangle =$tr$[ a \rho (t)] $. From the exact master equation
(\ref{eme}), it is easy to find that $\langle a(t) \rangle $ obeys
the equation of motion
\begin{align}
\langle \dot{a}(t) \rangle =-[ i\omega'_0(t) +\kappa(t)] \langle a
(t)\rangle =\frac{\dot{u}\left( t\right) }{%
u\left( t\right) }\langle a\left( t\right) \rangle .
\end{align}%
which has the exact solution:
\begin{equation}
\langle a (t) \rangle =u (t) \langle a (t_0) \rangle . \label{ma}
\end{equation}%
In other words, the temporal evolution of the cavity mode amplitude
is totally determined by $u (t)$, which indicates that $u(t)$ is the
propagating function characterizing the cavity field evolution, as
we have mentioned.

Another important physical observable is the total photon number
inside the cavity, which is defined by $n (t) =$tr$[ a^{\dag}a
\rho(t)]$. From the exact master equation, it is also easy to find
that
\begin{equation}
\dot{n} (t) =-2\kappa (t) n (t)
+\widetilde{\kappa} (t) . \label{diff-equ-nt}
\end{equation}%
On the other hand, Eq.~(\ref{coefs-b}) can be rewritten as
\begin{align}
\dot{v}(t) =-2\kappa(t)v (t) +\widetilde{\kappa} (t),
\end{align}%
with $-2\kappa(t)=[\dot{u}/u (t)+{\rm H.c.}]$. Combing these
equations together, we obtain the exact solution of $n(t)$ in terms
of $u(t) $ and $v(t) $:
\begin{equation}  \label{an}
n(t) =u(t)  n(t_0) u^{\ast}(t) +v(t) .
\end{equation}%
In fact, the above solution is a result of the correlated Green
function in nonequilibrium Green function theory \cite{Jin09101765}.
It contains two terms, the first term represents the temporal
evolution (usually a dissipation process) of the cavity field, due
to the coupling to the waveguide. The second term is a noise effect
induced by thermal fluctuation of the waveguide. Therefore,
Eq.~(\ref{an}) combines the dissipation and fluctuation dynamics
together to characterize the entire cavity dynamics. The dissipation
and fluctuation dynamics obeys the dissipation-fluctuation theorem,
as shown from the waveguide's time-correlation functions
(\ref{correla}). Since $v(t)$ is also determined by $u(t)$, as one
can see from Eq.~(\ref{v-e}), both the cavity mode amplitude and the
photon number inside the cavity are completely obtained by solving
the propagating function $u(t)$.

Furthermore, to see the coherence of the cavity field, we should
solve explicitly the reduced density operator.  This can be done
easily through the coherent state representation \cite{Tu08235311,
An07042127, Xiong1005}. Consider the cavity initially in a coherent
state,
\begin{equation}
\rho \left( t_{0}\right) =e^{-|\alpha _{0}| ^{2}} \left\vert \alpha
_{0}\rangle \langle \alpha _{0}\right\vert ,
\end{equation}%
it is not difficult to find  \cite{Xiong1005} that the reduced
density operator at arbitrary later time $t$ becomes
\begin{align}
\rho(t) = & \exp\Big\{\frac{|\alpha(t)|^2}{1+v(t)}\Big\}
\sum_{n=0}^{\infty } \frac{[ v(t)]^{n}}{[1+v(t)]^{n+1}} \notag \\
&~~~~~~~~~~~ \times \Big|\frac{\alpha(t)}{1+v(t)},n \Big\rangle
\Big\langle n, \frac{\alpha(t)}{1+v(t)}\Big| , \label{mcs}
\end{align}%
where $|\frac{\alpha(t)}{1+v(t)}, n\rangle \equiv
\exp[\frac{\alpha(t)} {1+v(t)}a^\dag]|n\rangle$ is a generalized
coherent state, and $\alpha(t)=u(t)\alpha_0$. It is interest to see
that Eq.~(\ref{mcs}) is indeed  a mixed state of generalized
coherent states $|\frac{\alpha(t)}{1+v(t)}, n \rangle$, in which the
photon number is given by
\begin{align}
n(t)=|u(t)\alpha_0|^2 + v(t) ,
\end{align}
as we expected.

Usually, $u(t)$ decays to zero due to the dissipation induced by the
coupling to the waveguide.  The corresponding reduced density
operator asymptotically becomes a thermally state with the
asymptotic photon number $n(t)=v(t \rightarrow \infty) \sim
\bar{n}(\omega_c, T)$. This solution shows precisely how the cavity
field loses its coherence (i.e. decoherence) due to the coupling to
the waveguide. This decoherence arises from the decay of the cavity
field amplitude $\alpha(t)=u(t)\alpha_0$ as well as the
thermal-fluctuation-induced noise effect manifested through the
correlation function $v(t)$, as shown in Eq.~(\ref{mcs}). The later
describes a process of randomly losing or gaining thermal energy
from the reservoir (here is the waveguide), upon the initial
temperature of the waveguide.

However, when the coupling between the cavity and the waveguide is
strong enough, $u(t)$ may not decay to zero, as we shall show
explicitly in the numerical calculation in the next section. Then
the reduced density operator remains as a mixed coherent state. On
the other hand, at zero-temperature limit $T=0$, we have
$\bar{n}\left(\omega ,T\right) =0$ so that $\widetilde{g} (\tau
-\tau{'}) =0$. As a result, we obtain $v (t) =0$. The reduced
density operator at zero temperature limit is given by
\begin{align}
\rho \left( t\right) |_{T=0} = e^{-|\alpha (t) |^{2}} |\alpha (t)
\rangle \langle \alpha (t)|.  \label{cst}
\end{align}%
In other words, the cavity can remain in a coherent state in the
zero temperature limit. These two features [$u(t)$ may not decay to
zero in the strong coupling regime and $v(t)=0$ at $T=0$] indicate
that enhancing the coupling between the cavity and the waveguide and
meantime lowing the initial temperature of the waveguide can
significantly reduce the cavity's decoherence effect in photonic
crystals.

\section{Numerical analysis of the exact non-Markovian dynamics}

\begin{figure}
\centering
\includegraphics[scale=0.6]{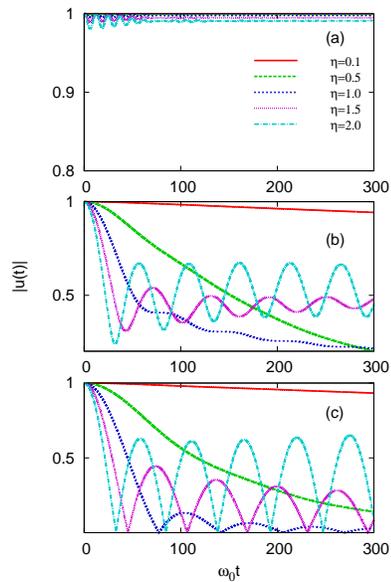}
\caption{(Color online) The exact solution of the scaled field
amplitude $|u(t)|$ of the microcavity in photonic crystals, coupled
to the waveguide with (a) $\omega_c=0.5 \omega_0$ (apart from the
waveguide band), (b) $\omega_c=1.025\omega_0$ (near the upper band
edge of the waveguide) and (c) $\omega_c=\omega_0$ (at the band
centre of the waveguide) from the weak coupling ($\eta < 0.7$) to
the strong coupling ($\eta
> 1.0$) regime.} \label{ut}
\end{figure}

\begin{figure}
\centering
\includegraphics[scale=0.7]{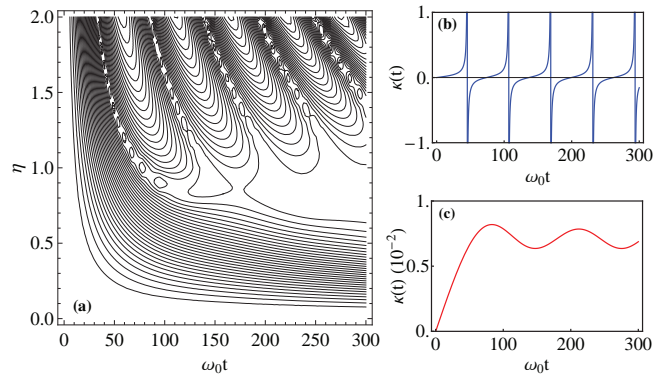}
\caption{(Color online) (a) A contour plot of the scaled cavity
field amplitude $|u(t)|$ by varying the time $t$ and the coupling
rate $\eta=\xi/\xi_0$, combined with other two plots for the decay
coefficient in the master equation, $\kappa(t)$, in (b) strong
coupling $\eta=1.5$ and (c) weak coupling $\eta=0.5$. The cavity
frequency is set to be the same as the resonator frequency of the
waveguide, $\omega_c=\omega_0$.} \label{kapper}
\end{figure}

\begin{figure}
\centering
\includegraphics[scale=0.7]{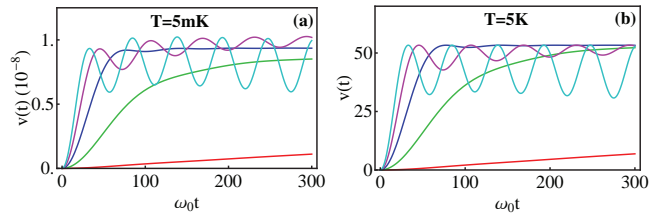}
\caption{(Color online) The temporal evolution of the
thermal-fluctuation-induced photon correlation function $v(t)$ in
the cavity coupled to the waveguide from weak coupling to strong
coupling regime with the initial temperature of the waveguide at (a)
$T=5$ mK, and (b) $T=5$ K. The curves of different colors with
different couplings are the same as in Fig.~\ref{ut}, here
$\omega_c=\omega_0$. } \label{vt}
\end{figure}

\begin{figure}
\centering
\includegraphics[scale=0.7]{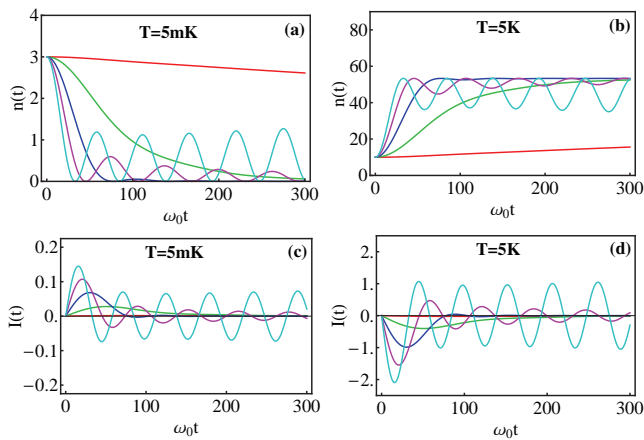}
\caption{(Color online) The temporal evolution of the photon number
$n(t)$ in the cavity and the photon current flowing in the
waveguide, by varying the coupling from weak ($\eta < 0.7$) to
strong ($\eta
> 1.0$) coupling regime with the initial temperature of the waveguide at (a) $T=5$ mK,
and (b) $T=5$ K. The curves of different colors with different
couplings are the same as in Fig.~\ref{ut}, here
$\omega_c=\omega_0$.} \label{nt}
\end{figure}

In this section, we will demonstrate the exact non-Markovian
dynamics of a microcavity coupled to a waveguide in photonic
crystals. In our calculation, based on the experiment in
Ref.~\cite{exp}, we take the frequency of the waveguide resonators
to be $\omega_{0} = 12.15$ GHz $=50.25 \mu$eV (in the unit
$\hbar=1$), and the coupling between the adjacent resonators to be
$\xi_{0} = 1.24 \mu$eV. The initial temperature of the waveguide is
set at $T = 5K$ so that $k_{B}T=430.75 \mu$eV $\approx 8.57
\omega_{0}$. The frequency of the single mode cavity $\omega_c$ and
the coupling $\xi$ between the cavity and the waveguide are tunable
parameters by changing the geometry of the cavity and the distance
between the cavity and the waveguide \cite{cavity with CROW-1,
PCcoupler, QDcoupledW}. With these experimental input parameters, we
numerically calculate the exact cavity dynamics with different
coupling strength for three different cavity frequency
configurations: i), the cavity coupled to the waveguide at the
waveguide band centre ($\omega_c = \omega_{0}$). ii), the cavity
coupled to the waveguide near the upper band edge ($\omega_{0} <
\omega_c < \omega_{0} + 2\xi_{0}$). iii), the cavity coupled apart
from the band of the waveguide ($\omega_c > \omega_{0}$). Detailed
numerical results are plotted in Figs.~\ref{ut}$-$\ref{nt}.

In Fig.~\ref{ut}, we show the exact solution of the scaled cavity
field amplitude, i.e. $|\langle a(t)\rangle/\langle
a(t_0)\rangle|=|u(t)|$ (see Eq.~(\ref{ma})), in different coupled
configuration from the weak coupling to strong coupling regime. For
$\omega$ lies outside the band of the waveguide, in both the weak
and strong coupling regimes, $|u(t)|$ remains unchanged beside a
short time very small oscillation at the beginning, see
Fig.~\ref{ut}(a). This result indicates that when its frequency lies
outside the band of the waveguide, the cavity effectively decouples
from the waveguide. However, when $\omega$ lies inside the band of
the waveguide, the time evolution behavior of the field amplitude is
totally different in different coupling regime, see
Fig.~\ref{ut}(b)-(c). In weak coupling regime (in terms of a
dimensionless coupling rate $\eta\equiv \frac{\xi}{\xi_0} < 0.7$),
$|u(t)|$ decays to zero monotonically, as a typical Markov process.
However, increasing the coupling such that $\eta > 1.0 $, after a
short time decay at the beginning, the field amplitude begins to
revives, and more than that, it keeps oscillating below an nonzero
value. This behavior shows that the cavity field no longer decays
monotonically in strong coupling regime, as a significant
non-Markovian memory effect. This effect becomes the strongest when
the cavity frequency matches the band centre of the waveguide, i.e.
$\omega_c=\omega_0$.

In Fig.~\ref{kapper}, we show a 3D plot of $|u(t)|$ varying in terms
of the coupling rate $\eta$ and the time $t$, where the critical
transition from the Markov to non-Markovian dynamics is manifested
with the critical coupling $\eta_c \simeq 0.7 \sim 1.0$. To
understand the underlying mechanism of this critical transition, we
also plot in Fig.~\ref{kapper} the decay coefficient in the master
equation (\ref{eme}), $\kappa (t) = -{\rm Re}[ \dot{u}(t) / u(t) ]$,
for different coupling configurations. The decay coefficient
$\kappa(t)$ dominates the dissipation behavior of the cavity field,
roughly given by the damping factor $\sim
e^{-\int_0^t\kappa(t')dt'}$. As one can see, in weak coupling
regime, after a short time increase, $\kappa (t)$ approaches to a
stationary positive value, see Fig.~\ref{kapper}(c). This leads to a
monotonic decay for the cavity field, i.e. a dissipation process.
However, in strong coupling regime, the behavior of $\kappa (t)$ is
totally different, it keeps oscillation in all the time between an
equal positive and negative bound value without approaching to zero,
see Fig.~\ref{kapper}(b). This oscillation process means that the
cavity dissipates energy to the waveguide and then fully regains it
back from the waveguide repeatedly. The overall effect of this
reviving process is that no energy dissipates into the waveguide. In
other words, the cavity dynamics becomes dissipationless in the
strong coupling regime. Thus, the critical transition from weak to
strong coupling regime reveals the transition from dissipation into
dissipationless processes for the cavity dynamics, as a
manifestation of the non-Markovian memory effect.

To see further the noise effect induced by thermal fluctuation in
the above non-Markovian process, we examine the temporal evolution
of the correlation function $v(t)$ given by (\ref{vt}). Physically,
Eq.~(\ref{an}) shows that if the cavity is initially empty, then
$v(t)$ is the average photon number inside the cavity, induced by
the thermal fluctuation of the waveguide. In Fig.~\ref{vt}(b), we
plot $v(t)$ with a few different coupling strength. As we see in the
weak coupling regime ($\eta < 0.7$), the exact $v(t)$ increases
monotonically and approaches to $\bar{n}(\omega_c, T)$ gradually.
However, in strong coupling regime ($\eta > 1.0$), the behavior of
$v(t)$ is qualitatively different from that in weak coupling case.
It increases much faster within a very short time in the beginning,
then keeps oscillation in a long time, in response to the
corresponding dissipationless oscillation of the cavity amplitude
$u(t)$.

To demonstrate explicitly the temperature dependence of this thermal
fluctuation effect, we plot $v(t)$ in Fig.~\ref{vt}(a) with a very
low temperature, $T=5$ mK. The value of $v(t)$ is reduced
dramatically ($ < 10^{-8}$ as shown in Fig.~\ref{vt}(a)). This
clearly shows that $v(t)$ characterizes the noise effect of the
thermal fluctuation from the waveguide. Lowing the initial
temperature of the waveguide can efficiently suppress the thermal
noise effect. Based on the analytical solution of the reduced
density matrix in the last section, if the cavity is initially in a
coherent state, and if the initial temperature of the waveguide is
low enough such that $v(t) \rightarrow 0$, the cavity state is given
by Eq.~(\ref{cst}) where $\alpha(t) = u(t) \alpha(0)$. As a result,
we can maintain well the cavity's coherence by enhancing coupling to
the waveguide such that the dissipation can also be suppressed.

To show the total non-Markovian memory effect distributing in the
dissipation and the noise processes, we examine the temporal
evolution of the photon number inside the cavity in
Fig.~\ref{nt}(a)-(b). The negative time derivation of the photon
number inside the cavity corresponds to the photon current flowed
into the waveguide, which is also shown in Fig.~\ref{nt}(c)-(d). The
photon current is an important quantity to characterize the
transmission of the waveguide. Eq.~(\ref{an}) shows that the total
photon number in the cavity consists of two sources: the evolution
of the initial photons in the cavity, and the thermal noise induced
photons. From Fig.~\ref{nt}(a)-(b) one see that in a relative high
temperature (above a few K), the thermal fluctuation, i.e. the
contribution from $v(t)$, dominates the photon number in the cavity,
where $v(t\rightarrow \infty) \sim \bar{n}(\omega_c, T)$ which is
about a few tens ($ \sim 50$ for $T=5$ K, as shown in
Fig.~\ref{vt}(b)) when $\omega_c$ is in the microwave region.
However, in a very low temperature, $v(t)$ approaches to zero ($v(t)
< 10^{-8}$ at $T=5$ mK, as shown in Fig.~\ref{vt}(a)). Then $n(t)$
is fully dominated by the evolution of the initial photon number in
the cavity, i.e. the first term $|u(t)|^2 n(t_0)$ in Eq.~(\ref{an}).

The difference of the time evolution of the photon number for the
weak and strong couplings is mainly shown in the long time behavior.
In weak coupling regime ($\eta < 0.7$), $n(t)$ approaches gradually
and monotonically to $\bar{n}(\omega_c, T)$. However, in the strong
coupling regime ($\eta > 1.0$), $n(t)$ quickly reaches to
$\bar{n}(\omega_c, T)$ and then oscillates around $\bar{n}(\omega_c,
T)$ due to the dissipationless oscillation of $u(t)$. In fact, the
dissipationless oscillation of $u(t)$ in strong coupling regime
indicates that the cavity field can produce subsequent pulses with a
small number of photons in the low temperature region. From
Fig.~\ref{nt}(a), we can see that the cavity can generate indeed
single photon pulses at $T=5$ mK when the coupling rate
$\eta=\xi/\xi_0=2$, namely the coupling between the cavity and the
waveguide is twice of the coupling between the adjacent resonators
in the waveguide, which is experimentally easy to realize.
Fig.~\ref{nt}(c)-(d) also plot the photon current in the waveguide,
which shows the corresponding oscillation associated with the
amplitude oscillation of the cavity field. Physically this result
indicates that the photon tunnels between cavity and the waveguide
repeatedly without loss of the coherence in the strong coupling
regime. Experimentally, one can directly measure the photon current
flowing over the waveguide to demonstrate these properties. These
properties may provide new applications for the microcavity in
photonic crystals.

From the above analysis, we find that when $\omega$ lies outside the
band of the waveguide, the cavity dynamics effectively decouples
from the waveguide. However, when $\omega$ locates inside the band
of the waveguide, the non-Markovian memory effect can qualitatively
change the dissipation as well as the noise dynamics of the cavity
field. In particular, the coupling between the microcavity and the
waveguide can manipulate well the dissipation behavior of the cavity
dynamics. Meanwhile, in the very low temperature limit, one can also
control efficiently the cavity coherence as well as the photon
number. Otherwise, the thermal fluctuation can induce non-negligible
noise effect.

\section{CONCLUSION}

In this paper, the exact non-Markovian dynamics of a  microcavity
coupled to a waveguide structure is studied. By solving the exact
master equation analytically, general solution of the density
operator as well as the cavity mode amplitude and the photon numbers
inside the cavity are obtained. We also examine the temporal
evolution of the cavity mode amplitude and the photon number
numerically in different coupling configurations. We show that
different cavity frequency and coupling between the cavity and the
waveguide would lead to totally different cavity dynamics. For the
frequency lies outside the band of the waveguide, the cavity becomes
isolated in both weak and strong coupling. When the cavity frequency
lies inside the band of the waveguide, the non-Markovian memory
effect qualitatively changes the amplitude damping behavior of the
cavity field as well as the thermal noise dynamics from the weak to
strong coupling regime. In particular, when the coupling strength
goes beyond a critical value, the cavity field becomes
dissipationless as a signature of strong non-Markovian memory
effect. The result also indicates that the cavity can maintain in a
coherent state with a very small photon number up to a single
photon, when it strongly couples to the waveguide at very low
temperature. The perfect transmission between the cavity and the
waveguide in photonic crystals is also more feasible in the strong
coupling regime at very low temperature. These properties can be
measured experimentally through the photon current flowing over the
waveguide in photonic crystals. We also hope that these properties
would provide further insights for the applications of the
microcavity in quantum optics.

\textit{Acknowledgements:} This work is supported by the National
Science Council of ROC under Contract No. NSC-96-2112-M-006-011-MY3.

\end{document}